\documentclass[11pt]{article}
\usepackage{amsmath}
\usepackage{amssymb}
\usepackage{graphicx}
\usepackage{color}
\usepackage{cancel}
\usepackage{fancybox}

\numberwithin{equation}{section}

\begin{document}
\begin{titlepage}
\begin{flushright}
%HRI-P-09-05-00X \\
KYUSHU-HET-118 \\
\end{flushright}

\begin{center}
\vspace*{15mm}
{\Large \bf The Role and Detectability of the Charm Contribution to Ultra High Energy Neutrino Fluxes}
\vspace{.5in}

Raj Gandhi${}^1$\footnote{raj@hri.res.in}, 
~Abhijit Samanta${}^1$\footnote{abhijit@hri.res.in},
~Atsushi Watanabe${}^{1,2}$\footnote{watanabe@hri.res.in/watanabe@higgs.phys.kyushu-u.ac.jp}
\vskip 0.5cm
{\small ${}^1${\it Harish-Chandra Research Institute, Chhatnag Road, Jhunsi,
Allahabad -211019, India}}\\
{\small ${}^2${\it Department of Physics, Kyushu University, Fukuoka 812-8581, 
Japan}}\\

\vspace{0.5cm}
{\small (May, 2009)}

\vspace{1cm}

\end{center}

\begin{abstract}\noindent%
It is widely believed that charm meson production and decay may play an 
important role in high energy astrophysical sources of neutrinos, 
especially those that are baryon-rich, providing an environment conducive 
to $pp$ interactions. 
Using slow-jet supernovae (SJS) as an example of such a source,
we study the detectability of high-energy neutrinos, paying particular 
attention to those produced from charmed-mesons. 
We highlight important distinguishing features in the ultra-high energy neutrino 
flux which would act as markers for the role of charm in the source.  
In particular, charm leads to significant event rates at higher energies, 
after the conventional ($\pi, K$) neutrino fluxes fall off.  
We calculate event rates both for a nearby single source and for diffuse SJS 
fluxes for an IceCube-like detector.  
By comparing muon event rates for the conventional and prompt fluxes 
in different energy bins, we demonstrate the striking energy dependence 
in the rates induced by the presence of charm. 
We also show that it leads to an energy dependant flux ratio of shower to muon 
events, providing an additional important diagnostic tool for the presence 
of prompt neutrinos.
Motivated by the infusion of high energy anti-electron neutrinos into the flux 
by charm decay, we also study the detectability of the Glashow resonance due to 
these sources.
\end{abstract} 

\end{titlepage}

\newpage
%%%%%%%%%%%%%%%%%%%%%%%%%%%%%%%%%%%%%%%%%%%%%%%%%%%%%%%%%%%%%%%%%%
\section{Introduction}
%%%%%%%%%%%%%%%%%%%%%%%%%%%%%%%%%%%%%%%%%%%%%%%%%%%%%%%%%%%%%%%%%%
Over the coming decade, the detection of cosmic neutrinos  is both
an exciting prospect and a challenge.
The considerable efforts deployed 
towards its realization are  made worthwhile by the potential to add to
our knowledge, both  of the astrophysics of high energy sources and of
physics beyond the standard model~\cite{AstroP}.
The extant and forthcoming ${\rm km}^3$ sized facilities~\cite{AMANDA,
IceCube,KM3NeT} may detect high-energy neutrinos from sources such as 
Active Galactic Nuclei (AGN) and Gamma Ray Bursts (GRB), 
heralding a new chapter in neutrino physics.
Specifically, if such neutrinos are observed, their intensity and spectrum
would provide key information towards understanding the dynamics of their 
source objects. 
They may also shed light on the origin of the observed cosmic rays.
Additionally, flavor ratios for the three types of neutrinos 
may provide insights into new physics beyond the standard model, e.g., 
neutrino decay, the presence of pseudo-Dirac mass states and 
CPT violation ~\cite{beyondSM,Beacom:2003eu,Barenboim:2003jm}.
The expected flux ratio $\Phi_e : \Phi_\mu : \Phi_\tau = 1: 2 : 0$ in many  
 sources is very sensitive to the existence of these  effects. 
If a significant deviation from the resulting democratic distribution 
$\Phi_e : \Phi_\mu : \Phi_\tau = 1: 1 :1$ expected at earth-based detectors 
(after oscillation effects are incorporated)  \cite{osc} is reconstructed from
the observations, it would be a clear signal of new physics.

An  unresolved issue which is expected to significantly affect the emitted 
and observed neutrino fluxes and their ratios is the role played by charm meson
production and subsequent decay. 
Such mesons are significantly short-lived compared to $\pi$ and $K$ mesons, 
which are copiously produced in the $p\gamma$ and $pp$ interactions which 
typify almost all such sources. 
This results in the charm meson neutrino spectrum in high energy astrophysical 
sources having qualitatively different features. 
Specifically, their decay is expected 
to lead to relatively flatter and elongated neutrino spectra compared to 
the sharply dropping contribution from $\pi$ and $K$ decays. 
This leads to an enhancement of higher energy neutrinos in sources 
where charm plays a role, as shown recently in~\cite{D}. Ratios of fluxes, 
which as stated above, are important diagnostics of new physics,  would also
be altered, since charm decay produces equal numbers of muon and electron 
neutrinos and antineutrinos.

While these effects may not be important in all high energy sources, 
they could significantly alter the observed neutrino fluxes for a class of 
sources which are characterized by proton rich jets, or, in general, 
an environment which renders $pp$ interactions important. It is our intent 
in this paper to  examine the observational consequences for ultra high energy 
detectors for this class of sources, where charm mesons are produced in 
significant numbers, along with $\pi$ and $K$. 
Our aim is to bring out general qualitative differences which would stand out 
as markers for the presence of charm production via $pp$ interactions 
in the source. If observed, this would also shed light on the issue of the 
presence and relative importance of $pp$ interactions versus $p\gamma$ 
interactions. 

As a specific case in point, to facilitate quantitative estimates of event 
rates  from which conclusions may be drawn, we choose as source slow-jet 
supernovae (SJS), which have been the focus of recent studies~\cite{D,RMW,AB}.
However, several important features related to observations which emerge from 
our calculations for SJS  may be generic to other types of sources whose jets 
are baryon rich. 

To this end, we calculate the event rates using the fluxes given in~\cite{
D,RMW,AB}. 
This is done both for a single SJS source and for the diffuse flux obtained 
by summing over sources. 
While it was shown in~\cite{RMW,AB} that a  nearby source 
(at around $3$ Mpc) is observable via its neutrinos from $\pi$ and $K$ decay, 
we show that the charm, or prompt component adds substantially 
to the neutrino event rate at energies significantly above those spanned 
by the conventional flux.
Additionally, we find that due to the charmed-meson contribution, 
the diffuse flux from SJS alone can rise above the atmospheric neutrino 
background, and may already be constrained by the AMANDA observations~\cite{
AMANDA}.
It must be noted, however, that flux estimates of high energy sources, 
though usually underpinned by reliable physics, have inherent uncertainties. 
This is more so in the case of charm contributions, with their added 
QCD-related uncertainty. 
The event numbers we compute are intended to serve as estimates, 
whereas the qualitative observational features which are pointed out here, 
distinguishing the presence of charm contributions,  may, 
in the long run, serve as more useful diagnostic tools as experiments 
accumulate data over the next decade. 

The plan of the paper is  as follows:
In Section 2, 
we briefly review the physics of  SJS  supplemented by charm meson production.
We discuss  the qualitative features of the flux and, 
for practical use, give its fitting functions.  
 We then calculate the event rates and  spectrum in an IceCube-like detector.
In Section 3, we obtain the diffuse flux and associated event rate 
by summing up all contributions from SJS  point sources.
Section 4 is devoted to a discussion of our results and  conclusions.

%%%%%%%%%%%%%%%%%%%%%%%%%%%%%%%%%%%%%%%%%%%%%%%%%%%%%%%%%%%%%%%%%%
\section{Neutrinos from Slow-Jet Supernovae}
%%%%%%%%%%%%%%%%%%%%%%%%%%%%%%%%%%%%%%%%%%%%%%%%%%%%%%%%%%%%%%%%%%
\subsection{The neutrino flux }
%%%%%%%%%%%%%%%%%%%%%%%%%%%%%%%%%%%%%%%%%%%%%%%%%%%%%%%%%%%%%%%%%%
Since the discovery of the connections between long duration GRB 
and supernovae~\cite{GRB-SNe},
it has been conjectured that a significant fraction of core-collapse 
supernovae (SNe) may be accompanied by mildly relativistic jets which   
do not break through the stellar envelope.
The choked jet may give rise to a radio after-glow without prompt
$\gamma$ ray signals, and such objects may be detectable primarily 
only via their neutrino emission.
The neutrino spectrum resulting from this scenario was first modeled by
Razzaque, M$\acute{\rm e}$sz$\acute{\rm a}$ros, and Waxman (RMW)~\cite{RMW} 
and extended by Ando and Beacom (AB)~\cite{AB}.
The essential difference between SJS and the related GRB jets is the mild 
gamma factor of the former,
$\Gamma \simeq 3$ and its baryon rich environment which makes $pp$ collisions 
 efficient.

An  extension of the work in~\cite{RMW,AB} has recently  
been discussed in~\cite{D}.
These authors calculate the neutrino flux from charm meson production and decay 
in the source, by solving the cascade equations which describe 
the propagation of protons and mesons.
The proton rich environment inside the slow jets is naturally conducive to charm
production via  $pp$ interactions. 
Daughter neutrinos from charmed mesons typically have energies higher than 
those resulting from conventional ones.
These mesons decay quickly, producing $\nu_{e,\mu} + \bar{\nu}_{e,\mu}$, 
before significant energy loss via interactions 
can occur, so that their contribution to the neutrino flux becomes dominant 
at $\gtrsim 10^{4.5} \,{\rm GeV}$.
This result is a significant change in the overall features of the emergent 
neutrino flux and, to an extent, in its flavor composition.

In general, meson spectra, whose  shapes   are  inherited by the daughter
neutrinos, are characterized by two power-law energy breaking scales.
The first breaking point $E^{(1)}$ is determined by the competition between 
decay  and hadronic cooling ( the latter characterized by, for instance, 
$\pi p$ and $K p$ interactions).
The second energy $E^{(2)}$ is determined by the transition between 
hadronic cooling and radiative cooling and marks the onset of the latter. 
Below the first breaking energy,  meson decay is dominant
and the spectrum is the same as that of the shock accelerated protons, 
$\sim E^{-2}$.
Between the first and the second scale, hadronic cooling is dominant,
and the spectrum gets steepen to $\sim E^{-3}$.
Above the second breaking energy, radiative cooling dominates,
and the spectrum falls as  $\sim E^{-4}$ up to the cutoff given by
the proton maximal energy.
For pions, the two breaking energies are close to each other
and relatively low; $E^{(1)} \sim 30 \, {\rm GeV}$ and
$E^{(2)} \sim 100 \, {\rm GeV}$.
Thus the pion flux retains an $E^{-4}$ power over TeV scales, 
up to the maximal cutoff energy.
For kaons, the relevant breaking energies are given by
$E^{(1)} \sim 0.2 \, {\rm TeV}$ and $E^{(2)} \sim 20\, {\rm TeV}$~\cite{AB}.

For the charmed mesons $D^\pm$ and $D^0$, due to their short lifetimes
and  large masses, the two scales are close to each other and
have relatively high values,  $E^{(1)} \sim E^{(2)} \sim 10^4 \, {\rm TeV}$
~\cite{D}.
Thus the  charmed flux  maintains a $E^{-2}$ power law all the way up
to the proton maximal energy, leading to a significantly flatter spectrum.

\begin{figure}[t]
\begin{center}
\scalebox{1.1}{
\includegraphics{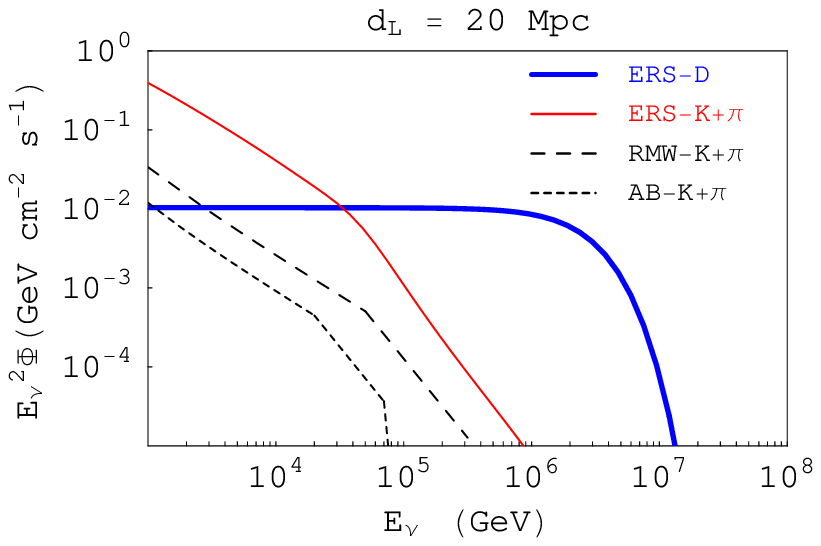}
}
\end{center}
\caption{
Muon neutrino and antineutrino fluxes from a source at 20 Mpc.
The thick(thin)-solid curve shows the $D^\pm$($K^\pm + \pi^\pm$) 
flux in~\cite{D}. 
The dashed and dotted curves are the $K^\pm + \pi^\pm$ fluxes
in~\cite{RMW} and~\cite{AB} respectively.
The fluxes are given in the earth-observer frame.
The effects of  neutrino oscillation are not incorporated in this set of curves.
}
\label{Fluxes1}
\end{figure}

In Fig.~\ref{Fluxes1}, we show various fluxes of muon neutrinos 
and antineutrinos at an Earth detector. The calculation uses RMW~\cite{RMW}, 
AB~\cite{AB} and ERS~\cite{D} source fluxes and propagates them to the earth  
for an assumed SJS source at a distance of $20$ Mpc. 
The fluxes are shown without neutrino oscillation effects, 
which we discuss in a  later sub-section. We note that the flux estimates 
for muon neutrinos from  $\pi$ and $K$ differ in the three models, 
while exhibiting similar spectral shapes. The differences, in part, 
result from differing treatments of the maximal proton cutoff energies.

For the ERS case, we also show the flux resulting from the incorporation 
of $D^\pm$-meson flux 
(The curve for $D^0$ is similar to that for  $D^\pm$). 
We note that the ERS-$D$ flux  becomes dominant above 
$E_\nu \sim 10^5 \, {\rm GeV}$.
This flux, after propagation over a distance $d_L$ from the source at 
redshift $z$ is well represented by the following fitting function,
\begin{eqnarray}
F_D  \,=\, \frac{ L_{\rm eff}^D }{2\pi d_L^2}\cdot
\frac{ (1 + z) }{E_\nu^2}\cdot
{\rm exp}\left[ 
-\left( \frac{(1 + z)E_\nu}{E_{\rm br}^D} \right)^{\beta_D} \, \right],
\label{Dfit}
\end{eqnarray}
where
\begin{eqnarray}
L_{\rm eff}^D &=& 2.6 \times 10^{50} \,\, {\rm GeV \cdot s^{-1}}, \\
E_{\rm br}^D  &=& 10^{6.5} \,\, {\rm GeV}, \\
\beta_D  &=& 1.35.
\end{eqnarray}
Similarly, for the conventional ERS ($\pi + K$ meson) flux, we find
\begin{eqnarray}
F_C  \,=\, \frac{ L_{\rm eff}^C }{2\pi d_L^2}\cdot
\frac{ (1 + z) }{E_\nu^2}\cdot
\left( \frac{(1 + z)E_\nu}{E_{\rm br}^C} \right)^{-\beta_C},
\label{Cfit}
\end{eqnarray}
where
\begin{eqnarray}
L_{\rm eff}^C &=& 3.1 \times 10^{50} \,\, {\rm GeV \cdot s^{-1}}, \\
E_{\rm br}^C  &=& 10^{4.5} \,\, {\rm GeV}, \\
\beta_C  &=&\left\{ \begin{array}{ccc}
 1.0  & {\rm for} & E_\nu < E_{\rm br}^C, \\
\vspace{-2.5mm}&& \\
 2.2  & {\rm for} & E_\nu > E_{\rm br}^C. \\
\end{array}
\right.
\end{eqnarray}
These expressions are useful not only for calculating the event numbers
but also to deduce the diffuse flux, both of which we discuss below.
The above fluxes do not accommodate the effect of
the meson accelerations by the internal shock~\cite{Koers:2007je}.
We note, however, that accounting for this acceleration and subsequent 
losses due to the increased lifetime and increased interactions would lower 
the ERS flux estimates, and the event rate predictions we make in this 
paper for them.

%%%%%%%%%%%%%%%%%%%%%%%%%%%%%%%%%%%%%%%%%%%%%%%%%%%%%%%%%%%%%%%%%%%%
\subsection{Neutrino oscillation effects}
%%%%%%%%%%%%%%%%%%%%%%%%%%%%%%%%%%%%%%%%%%%%%%%%%%%%%%%%%%%%%%%%%%%%
The neutrinos produced at a source can change flavors during
propagation to  earth.
Although their energies are very  high, the distance between a source 
and a detector is also large enough to average out the oscillatory behavior.
Writing the vacuum  transition probabilities between flavor eigenstates
$\alpha$ and $\beta$  in terms of the lepton mixing matrix
$V$ only, we have~\cite{osc}
\begin{eqnarray}
P_{\alpha \beta} &=& 
|V_{\alpha i}|^2|V_{ \beta i}|^2 \nonumber\\
 &\simeq& 
\begin{pmatrix}
1 - 2s& s &  s \\
s &  \frac{1}{2}(1 - s) &\frac{1}{2}(1 - s)  \\
s &  \frac{1}{2}(1 - s) &\frac{1}{2}(1 - s)  \\
\end{pmatrix},
\label{Pvac}
\end{eqnarray}
where $s \equiv \cos^2\theta_{12}\sin^2\theta_{12}$ and $\theta_{12}$ is
the solar angle in the standard  parameterization for $V$. 
The simple form of Eq.~(\ref{Pvac}) is a result of setting the ``reactor'' 
and  ``atmospheric'' mixing angles as $\theta_{13} = 0^\circ$ and 
$\theta_{23} = 45^\circ$ respectively.
As a consequence of this maximal $\mu$-$\tau$ mixing, the flux
ratio of $\Phi_\mu$ and $\Phi_\tau$ at earth must be 1. 
In addition, the standard ratio $\Phi_e : \Phi_\mu : \Phi_\tau = 1:2:0$ 
at a source means results in a democratic configuration 
$\Phi_e : \Phi_\mu : \Phi_\tau = 1:1:1$
at a detector, independent of the value of  $\theta_{12}$.

For SJS, the environment is baryon rich  and muons lose their
energy before decaying.
Thus, for the pion and kaon fluxes,  the initial configurations are 
$\Phi_e : \Phi_\mu : \Phi_\tau = 0:1:0$.
On the other hand, charmed mesons decay semi-leptonically
such that the initial ratio is $\Phi_e : \Phi_\mu : \Phi_\tau = 1:1:0$.
According to Eq.~(\ref{Pvac}), these initial fluxes at the source will be 
transmuted to 
$P \cdot (0,1,0)^{\rm T} = (s, (1-s)/2,(1-s)/2)^{\rm T} = 
(0.22,0.39,0.39)^{\rm T}$ and
$P \cdot (1,1,0)^{\rm T} = (1-s, (1+s)/2,(1+s)/2)^{\rm T} 
= (0.78,0.61,0.61)^{\rm T}$ 
at earth, where we have used $\theta_{12} = 34.5^\circ$ as a current best fit 
value \cite{data}. 
To summarize, we note the fluxes Eq.~(\ref{Dfit}) and Eq.~(\ref{Cfit}) after 
oscillation are given by
$F_C^e = s \cdot F_C$, $F_D^e =  (1-s)\cdot F_D$, 
$F_C^\mu = (1-s)/2 \cdot F_C$, and $F_D^\mu =  (1+s)/2\cdot F_D$, 
denoting the flavor species by upper subscripts.

\begin{figure}[b]
\begin{center}
\scalebox{1.1}{
\includegraphics{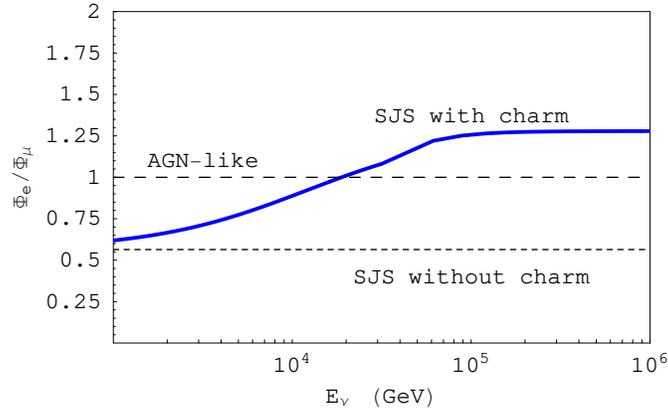}
}
\end{center}
\caption{The $e/\mu$ flavor ratio of the flux, as a function of 
the neutrino energy.
The solid curve shows the full contribution with SJS.
The dotted line is for SJS, but only with the conventional
component.
The dashed line represents a typical value for the sources which allow
full muon decay, such as AGN (without charm).}
\label{emuR}
\end{figure}

The difference of the initial flavor ratio between
the prompt and the conventional fluxes and their different spectral behavior
gives rise to an energy dependent  flavor ratio at  earth.
This is given by
\begin{eqnarray}
\frac{\Phi_e}{\Phi_\mu}=
2\frac{s\,\Phi_C + (1-s)\Phi_D}{(1-s)\Phi_C + (1+s)\Phi_D},
\label{ratio}
\end{eqnarray}
where $\Phi_{C,D}$ stand for generic charm and conventional fluxes.
The ratio is bounded as follows,
$2s/(1-s) \approx 0.56 < \Phi_e/\Phi_\mu < 2(1-s)/(1+s) \approx 1.28$
by the high and low-energy asymptotic behavior, marked by the dominance of 
the prompt and the conventional component respectively.
The energy dependence is most significant around the energy
at which the prompt and conventional components are equal.
We show the behavior of the ratio Eq.~(\ref{ratio}) by the solid curve
in Fig.~\ref{emuR}, by plotting the prompt and conventional 
flux Eq.~(\ref{Dfit}) and Eq.~(\ref{Cfit}) at $d_L = 20$ Mpc. 
Of course, this remains unaltered
when we compute the diffuse flux by integrating over the distribution
and luminosity of sources, since it depends only on the mixing matrix and 
the fact that the oscillation length is much shorter than the source distance. 

Fig.~\ref{emuR} highlights the importance of  detecting 
 both electron and muon flavored neutrinos, particularly
in the high-energy region above $50$ TeV.
The rise of the ratio above  unity at  high-energy  will be an
indication that charmed mesons contribute to  neutrino production.
In addition, the observation of the transition of the ratio around 
$1 \sim 100$ TeV
would indicate  that the source environments are baryon rich, to a degree 
that allows the charm component to rise above the conventional $\pi, K$ 
component and the muon to be damped. We also note the significant difference 
in the SJS without charm and
SJS with charm curves in the figure, i.e from a constant $0.5$ for the former, 
to an energy dependant transition from $0.5$ to $1.25$ for the latter. Thus, 
shower events would
register an increase from their low values compared to muon events at low 
energies, and rise above them at high energies.

We note that the canonical flavor ratio
$\Phi_e : \Phi_\mu : \Phi_\tau = 1:2:0$, the ratio
for muon damped sources, $\Phi_e : \Phi_\mu : \Phi_\tau = 0:1:0$
and that for semi-leptonic $D$ decay, $\Phi_e : \Phi_\mu : \Phi_\tau = 1:1:0$
should all be regarded as idealizations which can be used as first
approximations.
More precisely, there can be  small deviations from the above
standard ratios with some uncertainties~\cite{IFR}.
For example, it has been discussed that the muon damping cannot
be perfect and $\nu_e$ component consequently never can fall to zero~\cite{IFR}.
Inclusion of these effects will amount to finite thickness for
the curves presented in Fig.~\ref{emuR}.
However, qualitative behavior such as the transition of the ratio around 
$1 \sim 100$ TeV will not  change significantly and the curves in 
Fig.~\ref{emuR} still represent the essential qualitative physics of the 
charm effect.

%%%%%%%%%%%%%%%%%%%%%%%%%%%%%%%%%%%%%%%%%%%%%%%%%%%%%%%%%%%%%%%%%%%%
\subsection{Detection of high-energy neutrinos}
%%%%%%%%%%%%%%%%%%%%%%%%%%%%%%%%%%%%%%%%%%%%%%%%%%%%%%%%%%%%%%%%%%%%
%%%%%%%%%%%%%%%%%%%%%%%%%%%%%%%%%%%%%%%%%%%%%%%%%%%%%%%%%%%%%%%%%%%%%%%%
\begin{table}[t]
\begin{center}
\begin{tabular}{l|c|c|c|c|c|c}\hline\hline
&$10^{3\sim4}$&$10^{4\sim5}$&$10^{5\sim6}$& $10^{6\sim7}$&$10^{7\sim8}$&Total\\\hline
ERS-$D^{\pm,0}$&37&90&40&4&0&171 \\\hline
ERS-$K,\pi$&107&39&0.4&0&0&146 \\\hline 
AB-$K,\pi$&3&1&0&0&0&4\\\hline 
RMW-$K,\pi$&7&3&0&0&0&10 \\\hline 
\end{tabular}
\end{center}
\caption{The number of up-going $\nu_\mu N \to \mu X$ 
events from an SJS at $3$ Mpc.
The columns show the events in each decade of energy (in GeV)
and the total events.
An instrumented volume of $V = 1 \,(\rm km^3)$, 
and an energy threshold $E_\mu^{\rm min} = 10^3$ (GeV) is assumed.}
\end{table}
%%%%%%%%%%%%%%%%%%%%%%%%%%%%%%%%%%%%%%%%%%%%%%%%%%%%%%%%%%%%%%%%%%%%%%%%
We next compute the up and down-going  event rates for an IceCube-like detector 
for the fluxes considered in the previous sub-section, assuming  a point source 
at a distance of $3$ Mpc. While this has been previously done in \cite{RMW,AB}
for the conventional flux, our purpose here is to compare this contribution 
quantitatively with that of the charm mesons.

For up-going events, we incorporate  the attenuation
of neutrino flux by the earth matter, and also consider partially 
contained events
which enhance the effective volume of the detector significantly
for muon events.
The event rate is given by~\cite{UHEnu}
\begin{eqnarray}
{\rm Rate} =  A N_A 
\int \!\! dE_\nu \cdot \langle R(E_\mu^{\rm min},E_\nu) \rangle \cdot 
\sigma_{CC}(E_\nu)\cdot
S(E_\nu)\cdot F_{C,D}^\mu
\label{Rup}
\end{eqnarray}
where $A$ is the effective area of the detector, and $N_A$ is
Avogadro's number $N_A = 6.022 \times 10^{23} \,(\rm cm^{-3})$.
The muon range $\langle R(E_\mu^{\rm min},E_\nu) \rangle$ stands for 
the average distance traversed by a  muon  in earth matter.
Muons created by neutrinos with  energy $E_\nu$ attenuate
to the energy $E_\mu^{\rm min}$, after traveling 
$\langle R(E_\mu^{\rm min},E_\nu) \rangle$. 
The factor $S(E_\nu)$ represents the shadowing (i.e attenuation)
of the up-going neutrinos by the earth.

For down-going events, the event rate for the  $\nu N \to \mu X$
process is  given by
\begin{eqnarray}
{\rm Rate} =  N_A V_{\rm eff}
\int \!\! dE_\nu \cdot \sigma_{CC}(E_\nu)\cdot
F_{C,D}^\mu
\label{Rdown}
\end{eqnarray}
where $V_{\rm eff}$ is the effective volume of the detector,
assumed here to be  $ 1 \, (\rm km^3)$.
 $\sigma_{\rm CC}$ is the neutrino-nucleon  charged current cross section.

In addition to the $\nu N \to \mu X$ events, we have also calculated events for 
the Glashow resonance process, $\bar{\nu}_e e \to {\rm something}$,
which peaks around $E_\nu = 6.3 $ PeV, primarily because we expect an enhanced 
contribution from charm at the higher energies.
The event rate for down-moving neutrinos from this channel is given by
\begin{eqnarray}
{\rm Rate} = \frac{10}{18} N_A V_{\rm eff}
\int \!\!dE_\nu \cdot \sigma_{\bar{\nu}_e e}(E_\nu)\cdot
\frac{1}{2}F_{C,D}^e
\label{RGlashow}
\end{eqnarray}

\begin{figure}[t]
\begin{center}
\scalebox{1.1}
{
\includegraphics{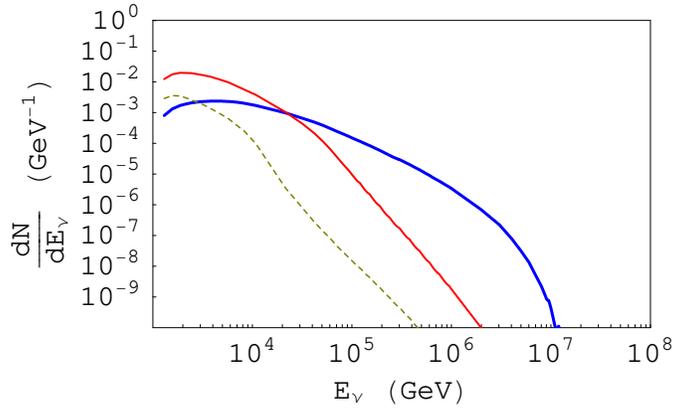}
}
\end{center}
\caption{Event spectrum of up-going $\nu N \to \mu^- X$ 
with a source at $d_L = 3$ Mpc for ERS flux~\cite{D}. 
The blue (thick solid), red (thin solid) and the 
dashed curve shows $D^{\pm}$, $K$ and $\pi$ contribution
respectively.
}
\label{f1}
\end{figure}

\begin{figure}[b]
\begin{center}
\scalebox{1.1}
{
\includegraphics{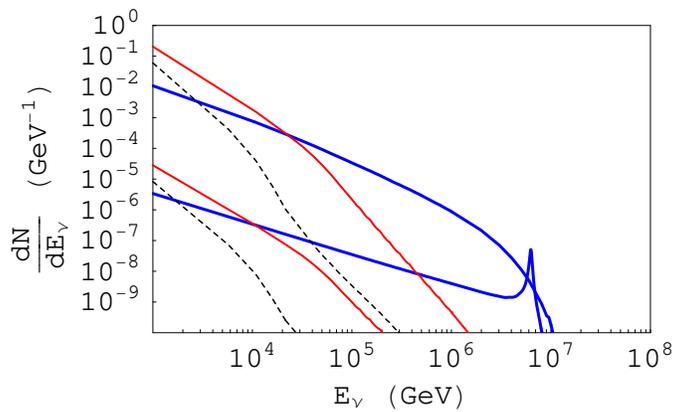}
}
\end{center}
\caption{Same as Fig.~\ref{f1} but for down-going events.
The lower curve in each case  is for 
$\bar{\nu}_e e \to \bar{\nu}_\mu \mu$ channel.
}
\label{f2}
\end{figure}

Our results for all the various  event spectra are shown 
in Fig.~\ref{f1}(\ref{f2}) for up(down)-going events. 
Clearly, the charm contribution dominates above $\sim$ $50$ TeV, 
and its contribution is significant even around $500-1000$ TeV, 
where the $\pi$ and $K$ contributions to the event rate  are negligible.
Table~1(2) shows the integrated number of events
for each energy decade with up(down)-going muon events.
First, we note that the ERS flux, even for the conventional case, yields rates 
which are about 20-30 times higher than the AB or RMW fluxes, due to an 
enhanced $pp$ contribution in this model. 
The charm contribution for ERS is a full third of the integrated rate for the 
conventional flux. Importantly, it manifests itself in energy regions where 
the conventional contribution is low, i.e., in the band $10^4-10^6$ GeV. 
This both enhances its prospects for detection and acts as a distinctive 
signature for charm.
Finally, we note that since the typical duration of the burst is a few seconds, 
the background from the  atmospheric flux is small.

In Fig.~\ref{f2}, we show the event spectrum for 
$\bar{\nu}_e e \to \bar{\nu}_\mu \mu$ channel as well.
From the figure, we can clearly see the resonant behavior 
at $6.3$ PeV.
The number of events for this mode is, with the ERS-$D$ flux, $0.06$ 
in the  $10^{5\sim6}$ GeV bin. While this contribution is $\sim 50$ \% of its  
total event tally over  the full energy range of $10^{3\sim8}$ GeV, it is 
unfortunately too tiny to be detectable.

%%%%%%%%%%%%%%%%%%%%%%%%%%%%%%%%%%%%%%%%%%%%%%%%%%%%%%%%%%%%%%%%%%%%%%%%
\begin{table}[t]
\begin{center}
\begin{tabular}{l|c|c|c|c|c|c}\hline\hline
&$10^{3\sim4}$&$10^{4\sim5}$&$10^{5\sim6}$& $10^{6\sim7}$&$10^{7\sim8}$&Total\\\hline
ERS-$D^{\pm,0}$&44&26&9&1.3&0&80 \\\hline
ERS-$K,\pi$&228&13&0.1&0&0&241 \\\hline 
AB-$K,\pi$&7&0.3&0&0&0&7\\\hline 
RMW-$K,\pi$&17&1&0&0&0&18 \\\hline 
\end{tabular}
\end{center}
\caption{Same as Table 1, but for down-going $\nu_\mu N \to \mu X$ events.}
\end{table}
%%%%%%%%%%%%%%%%%%%%%%%%%%%%%%%%%%%%%%%%%%%%%%%%%%%%%%%%%%%%%%%%%%%%%%%%

%%%%%%%%%%%%%%%%%%%%%%%%%%%%%%%%%%%%%%%%%%%%%%%%%%%%%%%%%%%%%%%%%%%%
\section{ The Diffuse neutrino flux from SJS}
%%%%%%%%%%%%%%%%%%%%%%%%%%%%%%%%%%%%%%%%%%%%%%%%%%%%%%%%%%%%%%%%%%%%
\subsection{The diffuse neutrino flux}
%%%%%%%%%%%%%%%%%%%%%%%%%%%%%%%%%%%%%%%%%%%%%%%%%%%%%
Prior to presenting our results for the events from the diffuse SJS fluxes, 
we note that charm is likely to contribute to the atmospheric flux as well.
This has been discussed in detail in, for example,
in~\cite{ATMcharm}. We have ascertained that  its contribution is low and 
does not constitute a significant background to the SJS  flux. 

The diffuse neutrino flux is obtained by integrating the point source
flux over all SNe, weighting the flux by a rate 
$\frac{d^2 N_{\rm sn}}{dt d\Omega}$, the number of SNe events
per unit time per solid angle covering the earth sky.
The distribution $\frac{d^2 N_{\rm sn}}{dt d\Omega}$ is given by~\cite{RMW}
\begin{eqnarray}
\frac{d^2 N_{\rm sn}}{dt d\Omega} = 
\frac{\dot{n}_{\rm sn}(z) d_L^2 c}{(1 + z)^2}\left| \frac{dt}{dz} \right|,
\end{eqnarray}
where the SNe rate $\dot{n}_{\rm sn}(z)$ is 
\begin{eqnarray}
\dot{n}_{\rm sn}(z)
&=& 0.017 \,\frac{0.32 e^{3.4z}}{e^{3.8z} + 45} \times 10^{-81} \quad 
{\rm (s^{-1} \cdot cm^{-3})}.
\end{eqnarray}
The cosmic time $t$ and $z$ is related as
\begin{eqnarray}
\left| \frac{dt}{dz} \right| \,=\,
\frac{1}{H_0}\frac{1}{1 + z}\frac{1}{\sqrt{
(1 + \Omega_m z)(1 + z)^2 - \Omega_\Lambda (2z + z^2) }},
\end{eqnarray}
where $\Omega_m = 0.3$ and $\Omega_\Lambda = 0.7$ within the standard
cosmology.
With these ingredients at hand, 
the diffuse neutrino flux is given by (e.g. for charm flux)
\begin{eqnarray}
\Phi^{\rm diff}_{D} &\,=\,&
\frac{\xi_{\rm sn}}{2\Gamma^2}
\int_0^\infty \!\! dz \cdot
\frac{d^2 N_{\rm sn}}{dt d\Omega}\cdot t_jF_D^\mu \nonumber\\
&\,=\,&
\frac{\xi_{\rm sn}}{4\pi \Gamma^2}\cdot
\frac{c L_{\rm eff}^D t_j }{E_\nu^2}
\int_0^\infty \!\! dz \cdot
\frac{\dot{n}_{\rm sn}(z)}{(1 + z)}
\left| \frac{dt}{dz} \right|
{\rm exp}\left[ 
-\left( \frac{(1 + z)E_\nu}{E_{\rm br}^D} \right)^{\beta_D}  \, \right],
\label{Dflux}
\end{eqnarray}
where $t_j$ is the jet duration of $\simeq 10 \,{\rm s}$, and
$\xi_{\rm sn}$ represents a rate of SNe with an accompanying jet.
The probability that the jet points to the earth is assumed to be
$1/2\Gamma^2$.

\begin{figure}[t]
\begin{center}
\scalebox{1.2}{
\includegraphics{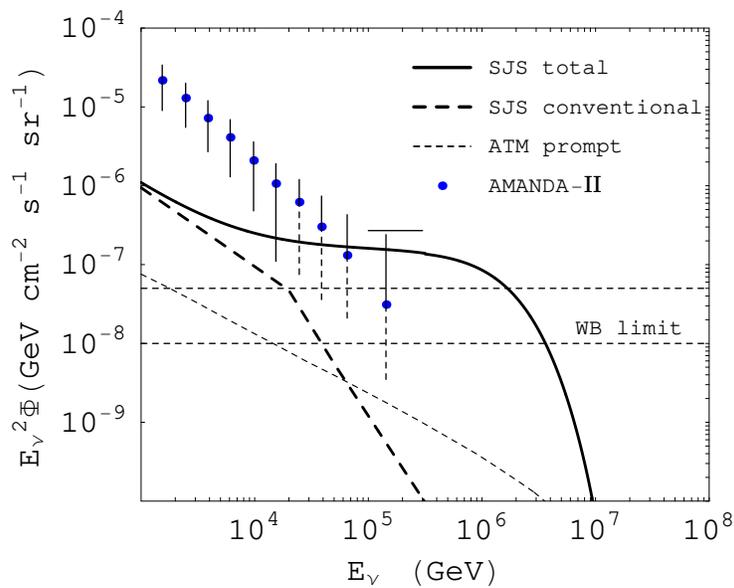}
}
\end{center}
\caption{The diffuse $\nu_\mu + \bar{\nu}_\mu$ flux from 
slow-jet supernovae, 
compared with the reconstructed atmospheric neutrino flux by 
AMANDA 2000 year data \cite{AMANDA}.
The thick plain (dashed) curve shows the total (conventional) 
flux for SJS, and the thin dashed curve is the prompt flux from
the atmosphere \cite{ATMcharm}.
The horizontal dashed lines shows the WB limit 
$\sim 1-5 \times 10^{-8} \,\, {\rm GeV \cdot cm^{-2} \cdot s^{-1} 
\cdot sr^{-1}}$~\cite{WBb}.
}
\label{Diffuseplot}
\end{figure}

In Fig.~\ref{Diffuseplot}, we show $\nu_\mu + \bar{\nu}_\mu$ 
diffuse fluxes.
From the figure, we can see that the total contribution for the ERS flux  
already begins to breach AMANDA data, and hence the charm contribution is 
likely to be both tested and constrained over the coming years. 
As already pointed out, the range over which it is dominant is largely 
 different from that spanned by the conventional contribution from 
$\pi$ and $K$ mesons, making it observationally distinct.

Table 3 shows the number of the events per year per steradian for
each flux.
In Fig.~\ref{Diffuseplot} and Table 3, we have, somewhat unrealistically, 
endowed all supernovae with  slow jets of $\Gamma = 3$, and thus 
taken $\xi_{\rm sn} = 1$.
For the ERS fluxes, the diffuse rates are quite high and certainly observable,
especially in the higher background free energy bins. Roughly, a total of 200
up-going and 180 down-going events are expected per year\footnote{ We note 
that the Waxman-Bahcall bound~\cite{WBb}, which appears to be breached, 
does not apply to such sources since they do not contribute to 
the cosmic-ray flux}.
A more realistic value of $\xi_{\rm sn}$ would lower our rates by an appropriate 
factor.

Finally, we present simple fits to the flux curves
for the muon neutrinos shown in Fig.~\ref{Diffuseplot}.
The fluxes are well represented by
\begin{eqnarray}
y = ax^b + c e^{dx},
\label{fit}
\end{eqnarray}
where $y = \log_{10}(\Phi E_\nu^2), x = \log_{10}(E_\nu)$. 
The diffuse flux $\Phi$ and the neutrino energy $E_\nu$ are
in the same units as in Fig.~\ref{Diffuseplot}.
Using Eq.~(\ref{fit}), the neutrino flux for each case, (i) SJS conventional 
(ii) SJS $D$-meson and (iii) SJS total can be fitted and the  sets of
the best-fit  parameters, $a,b,c$, and $d$ are listed in Table \ref{t:fit}.

%%%%%%%%%%%%%%%%%%%%%%%%%%%%%%%%%%%%%%%%%%%%%%%%%%%%%%%%%%%%%%%%%%%%%%%%
\begin{table}[t]
\begin{center}
\begin{tabular}{l|c|c|c|c|c|c}\hline\hline
&$10^{3\sim4}$&$10^{4\sim5}$&$10^{5\sim6}$& $10^{6\sim7}$&$10^{7\sim8}$&Total\\\hline
ERS-$D^{\pm,0}$&34&82&36&4&0&156 \\\hline
ERS-$K,\pi$&45&16&0&0&0&61 \\\hline 
AB-$K,\pi$&0.7&0.2&0&0&0&1\\\hline 
RMW-$K,\pi$&2&0.5&0&0&0&3 \\\hline 
\end{tabular}
\end{center}
\caption{The number of up-going muon events per year per steradian 
with the diffuse prompt and conventional fluxes.
The instrumented volume of the detector is taken as $V = 1 \,(\rm km^3)$;
the effective area $A = 1\, (\rm km^2)$ for up-going and 
the effective volume $V_{\rm eff} = 1\,(\rm km^3)$ for down-going.  
We are taking the energy threshold as $E_\mu^{\rm min} = 10^3$ (GeV).}
\end{table}
%%%%%%%%%%%%%%%%%%%%%%%%%%%%%%%%%%%%%%%%%%%%%%%%%%%%%%%%%%%%%%%%%%%%%%%%
%%%%%%%%%%%%%%%%%%%%%%%%%%%%%%%%%%%%%%%%%%%%%%%%%%%%%%%%%%%%%%%%%%%%%%%%
\begin{table}[t]
\begin{center}
\begin{tabular}{l|c|c|c|c|c|c}\hline\hline
&$10^{3\sim4}$&$10^{4\sim5}$&$10^{5\sim6}$& $10^{6\sim7}$&$10^{7\sim8}$&Total\\\hline
ERS-$D^{\pm,0}$&40&24&8&1&0&73 \\\hline
ERS-$K,\pi$&98&5&0&0&0&103 \\\hline 
AB-$K,\pi$&1&0&0&0&0&1\\\hline 
RMW-$K,\pi$&4&0.2&0&0&0&4 \\\hline 
\end{tabular}
\end{center}
\caption{
Same as Table 3, but for down-going muon events.
}
\end{table}
%%%%%%%%%%%%%%%%%%%%%%%%%%%%%%%%%%%%%%%%%%%%%%%%%%%%%%%%%%%%%%%%%%%%%%%%
{
\begin{table}[t]
\begin{center}
\begin{tabular}{l|c|c|c|c}\hline\hline
 & $a$ & $b$ & $c$ & $d$ \\\hline
(i) SJS conventional & $-0.86203$ &1.41904 &$-25.73761$ &$-0.84254$\\\hline
(ii) SJS $D$-meson &$-6.20429$&0.06759&$-6.8552\times 10^{-9}$&2.83302 \\\hline
(iii) SJS total &$-5.17119$&0.16516&$-2.1531 \times 10^{-8}$&2.68061 \\\hline
\end{tabular}
\end{center}
\caption{Best-fitted values of the parameters $a,~b,~c,~d$ for different SJS fluxes.}
\label{t:fit}
\end{table}

}
%%%%%%%%%%%%%%%%%%%%%%%%%%%%%%%%%%%%%%%%%%%%%%%%%%%%%%%%%%%%%%%%%%%%%%%%
\section{Summary and Conclusion}
%%%%%%%%%%%%%%%%%%%%%%%%%%%%%%%%%%%%%%%%%%%%%%%%%%%%%%%%%%%%%%%%%%%%%%%%
The production and decay of charmed mesons can play an important role 
for the detection of high-energy cosmic neutrinos.
Slow-jet supernovae provide both an excellent example and a test case 
in which the charm effect can be significant due to the baryon rich 
environment, 
leading to the rise of $pp$ contributions to the total neutrino rate.
This enhances the detectability of the SJS neutrinos
not only with the point sources but also with the diffuse flux.

The charm contribution appears to open up the possibility of detecting SJS via 
their diffuse flux, using the ERS flux calculations.
While the absolute intensity of the flux should be regarded with caution
 at the  present stage, the elongated shape of the total flux offers a shot at 
detection, provided $pp$ interactions play a significant role, and if the 
the jet parameters (and the faction $\xi_{\rm sn}$ ) are favorable.

By virtue of the semileptonic decay of the charmed mesons, the flavor ratio 
$\Phi_e/\Phi_\mu$ must be $\simeq 2(1 - \sin^2\theta_{12})$, 
which is essentially different from the typical AGN or GRB case.
Furthermore, the energy dependence of $\Phi_e/\Phi_\mu$ also carries
potentially important information.
The ratio shifts occurs around the crossing point $E_{\rm cross}$
where the $D$-meson flux becomes equal to the kaon flux.
The crossing energy is written as $E_{\rm cross} \sim (f_K/f_D)E_K^{(1)}
\approx 200 E_K^{(1)}$, where $f_K(f_D) \simeq 0.06(3\times 10^{-4})$
are the kaon and the charmed-meson multiplicities, and $E_K^{(1)}$ is
the first breaking energy for kaon.
The breaking energy $E_K^{(1)}$ is very sensitive
to the bulk Lorentz factor $\Gamma$ such that $E_K^{(1)} \propto 
\Gamma^{7\div 5}$ depending on the assumption on the jet-opening angle~\cite{
AB}.

To summarize, charmed mesons can play an important role in a certain class
of high energy sources in general and some  supernovae in particular, 
where jets 
are mildly relativistic and environments baryon-rich. Detection of the 
charm related neutrino flux may be possible with current and near-future 
detector sensitivities, and  can offer a valuable handle on several important 
open questions related to the dynamics of such sources.

%%%%%%%%%%%%%%%%%%%%%%%%%%%%%%%%%%%%%%%%%%%%%%%%%%%%%%%%%%%%%%%%%%%%%%%%%%%
\vspace{2mm}
\subsection*{Acknowledgments}
The authors acknowledge support from the Neutrino Project under the
XIth plan of Harish-Chandra Research Institute,
and thank Mary Hall Reno and Rikard Enberg for useful discussions.
%%%%%%%%%%%%%%%%%%%%%%%%%%%%%%%%%%%%%%%%%%%%%%%%%%%%%%%%%%%%%%%%%%%%%%%%%%%

%%%%%%%%%%%%%%%%%%%%%%%%%%%%%%%%%%%%%%%%%%%%%%%%%%%%%%%%%%%%%%%%%% 

\newpage

%%%%%%%%%%%%%%%%%%%%%%%%%%%%%%%%%%%%%%%%%%%%%%%%%%%%%%%%%%%%%%%%%% 
%%%%%%%%%%%%%%%%%%%%%%%%%%%%%%%%%%%%%%%%%%%%%%%%%%%%%%%%%%%%%%%%%% 
%%%%%%%%%%%%%%%%%%%%%%%%%%%%%%%%%%%%%%%%%%%%%%%%%%%%%%%%%%%%%%%%%% 
%%%%%%%%%%%%%%%%%%%%%%%%%%%%%%%%%%%%%%%%%%%%%%%%%%%%%%%%%%%%%%%%%% 
%%%%%%%%%%%%%%%%%%%%%%%%%%%%%%%%%%%%%%%%%%%%%%%%%%%%%%%%%%%%%%%%%% 
%%%%%%%%%%%%%%%%%%%%%%%%%%%%%%%%%%%%%%%%%%%%%%%%%%%%%%%%%%%%%%%%%% 
%%%%%%%%%%%%%%%%%%%%%%%%%%%%%%%%%%%%%%%%%%%%%%%%%%%%%%%%%%%%%%%%%% 
%%%%%%%%%%%%%%%%%%%%%%%%%%%%%%%%%%%%%%%%%%%%%%%%%%%%%%%%%%%%%%%%%% 
%%%%%%%%%%%%%%%%%%%%%%%%%%%%%%%%%%%%%%%%%%%%%%%%%%%%%%%%%%%%%%%%%% 
%%%%%%%%%%%%%%%%%%%%%%%%%%%%%%%%%%%%%%%%%%%%%%%%%%%%%%%%%%%%%%%%%% 

\end{document}